\documentstyle{article}
\input math_macros
\font\tenbf=cmbx10
\font\tenrm=cmr10
\font\tenit=cmti10
\font\elevenbf=cmbx10 scaled\magstep 1
\font\elevenrm=cmr10 scaled\magstep 1
 1

\font\ninerm=cmr9

\begin{document}
\begin{center}{{\tenbf MINIMAL SO(10) GRAND UNIFICATION: PREDICTIONS\\
               \vglue 10pt
FOR PROTON DECAY AND  NEUTRINO MASSES AND MIXINGS \\}
\vglue 10pt
{\tenbf {R.N. MOHAPATRA}\footnote{\ninerm\baselineskip=11pt
{ Invited} {talk presented at the workshop on "New Physics with New
Experiments" held in Kaimierz, Poland, May, 1993.}}
 \\}
{\tenit Department of Physics, University of Maryland,
College Park, MD 20742, USA\\}
\vglue 0.8cm
{\tenrm ABSTRACT}}
\end{center}
\vglue 0.3cm
{\rightskip=3pc
 \leftskip=3pc
 \tenrm\baselineskip=12pt
 \noindent

Prospects for SO(10) as a minimal grand unification group have
recently been heightened
by several considerations such as the MSW resolution
of the solar neutrino puzzle, baryogenesis, possibility
for understanding fermion masses etc. I review the
present status of the minimal SO(10) models with special
emphasis on the predictions for proton lifetime and predictions
for neutrino masses for the  non-supersymmetric case and
discuss some preliminary results for the supersymmetric case.
It was generally believed that  minimal SO(10) models predict
wrong mass relations between the charged fermions of the first
and second generations; furthermore, while the smallness of
the neutrino masses  in these models arises from the see-saw
mechanism, it used to be thought that detailed predictions for
neutrino masses and mixings require further adhoc assumptions.
In this talk, I report some recent work with K.S.Babu, where we
discovered that the minimal SO(10) model, both with and without SUSY,
has in it a built-in mechanism that not only corrects the bad mass
relations between the charged fermions but at the same time allows
a complete prediction for the masses and mixings in the neutrino
sector.  We define our minimal model as the one that consists
of the smallest set of Higgs multiplets that are needed for
gauge symmetry breaking. Our result is based on
 the hypothesis that the complex {\bf 10} of Higgs bosons
has only a single coupling to the fermions. This hypothesis is
guaranteed in supersymmetric models and in non-SUSY models that
obey a softly broken Peccei-Quinn symmetry.

\vglue 0.6cm}
\baselineskip=14pt
\elevenrm

\def\Tilde#1{\widetilde {#1}}

\noindent{\elevenbf 1. Introduction:}

At the present time ,the only experiments where there is some hint
of new physics beyond the standard model are the ones which have
detected the neutrinos emitted from the solar core[1].
The deficit of solar neutrinos reported in these experiments from
Homestake, Kamiokande, SAGE, and GALLEX[2] if confirmed in the other
proposed experiments such as the BOREXINO[3] and SNO[4]
will confirm that
the neutrinos have masses and mixings very much like the quarks.
The observed deficit in the present experiments
 can be explained in terms of neutrino oscillations in two
different ways: (i) long wave length vacuum oscillation[5], and (ii)
resonant matter oscillation (the Mikheyev-
Smirnov-Wolfenstein (MSW) effect[6]).
Assuming a two-flavor $\nu_e-\nu_\mu$ oscillation,
in the former case, the neutrino masses
and mixing angle should satisfy
$\Delta m^2  \sim
10^{-10}~eV^2$ and sin$^22\theta_{e\mu} \simeq (0.75~ {\rm to}~1)$.  In case
of MSW there are two allowed windows that fit
all of the experimental data[7]:
(a) the small mixing angle non--adiabatic
solution, which requires $\Delta m^2 \simeq (0.3~ {\rm to}~1.2)\times
10^{-5}~eV^2$ and ${\rm sin}^22\theta_{e\mu} \simeq (0.4~{\rm to}~1.5)
\times 10^{-2}$, and
(b) the large angle solution with $\Delta m^2\simeq (0.3~{\rm to}~5)
\times 10^{-5}~eV^2$
and ${\rm sin}^22\theta_{e\mu} \simeq (0.5~{\rm to}~0.9)$.
In all these cases, barring an unlikely scenario of near mass degeneracy among
neutrinos, either $\nu_\mu$ or $\nu_\tau$ should have mass in the
$(10^{-5}~{\rm to} ~10^{-3})~eV$ range. Specifically for the MSW
resolution, the mass has to be of order $10^{-3} eV$.

There are two other indications, though more controversial, for
a non-vanishing neutrino mass; i) the apparent deficit of GeV muon
neutrinos in the cosmic rays that originate from the decay of kaons
and pions[8] in the atmosphere;
 and ii) the need for a hot component in the dark matter
of the universe[9]. The first effect would be an indication of a
possible oscillation between the $\nu_{\mu}$ and $\nu_{\tau}$, again
indicating the existence of a non-vanishing mass for the neutrinos.
The simplest candidate for the hot component of the dark matter is
a neutrino wth mass in the few electron volt range.
The controversy surrounding the first effect has to do with the
uncertainties in the atmospheric muon neutrino fluxes[10],
whereas that with the second is due to possible effects of a
cosmological constant and other cosmological inputs that go into
the discussion of structure formation. In fact if all these three
effects are to be taken together, the resulting form of the neutrino
mass matrix becomes very seriously constrained[11] and one has
to invoke specific kinds of family symmetries to understand them.
Our goal in this article will be to see to what extent the present
observations motivate the serious consideration of SO(10) as a
grand unification symmetry and how much of the data can be understood
in this framework.

A natural explanation for the origin of such tiny neutrino
masses is the see--saw mechanism[12], which is based on a mass
matrix of the following form:
\renewcommand\arraystretch{1.1}
\begin{equation} M= \left( \begin{array}{cc}
  0  &  m_{D} \\
m^T_{D} & M_N   \end{array}     \right)
\end{equation}

Here ,the $m_D$ and $M_N$ are $3\times 3$ matrices denoting the
Dirac and the Majorana masses involving the left and the right-handed
neutrinos. The diagonalization of the see-saw matrix leads to
the light neutrino masses of the following form:
\begin{equation}
      M_{\nu}\simeq {{m^2_D}\over {M_N}}
\end{equation}

The mass matrix $M_N$ corresponds to the scale of $B-L$ breaking and
represents physics beyond the standard model. It therefore could be
large whereas $m_D$ is characteristic of the electroweak scale and
is in the GeV range, thus explaining the smallness of the neutrino masses.
  The solar neutrino
puzzle indicates that the $B-L$ scale is in the
$(10^{12}-10^{16})~GeV$ range.

All of the observations above, viz., non--zero neutrino masses, the see--saw
mechanism, and a high $B-L$ scale, fit rather naturally in grand
unified models based on the gauge group $SO(10)$.
In its non-supersymmetric version, experimental
constraints from proton life-time and the weak mixing angle sin$^2\theta_W$
require that $SO(10)$ breaks not directly into the standard model, but at
least in two steps.  In a two-step breaking scheme, the left--right
symmetric
intermediate scale is around $10^{12}$ GeV[13].
In supersymmetric $SO(10)$ there is no need for an
intermediate scale, $SO(10)$ can break directly to the
standard model at around $10^{16}$ GeV.

   There is also another compelling cosmological reason for the $SO(10)$
model which is absent in GUT models ,which has to do with the generation
of cosmological baryon asymmetry. It has been pointed out that, in $SO(10)$
models with the see-saw mechanism ,the right-handed neutrino is very heavy
and has a mass near the $B-L$ breaking scale of $10^{12}$ GeV or so.
Therefore,its decay can generate a lepton asymmetry  when the temperature
of the Universe is about $10^{11}$ GeV or so. This lepton asymmetry
gets subsequently converted to the baryon asymmetry due the sphaleron
effects at around $T\simeq 300$GeV[14]. In investigating the survival
of the lepton asymmetry from this high temperature down to the
electro-weak phase transition temperature , one must make sure that
the lepton violating interactions are out of equilibrium. This is
guaranteed only if the $B-L$ symmetry scale is bigger than $10^{11}$
or so. This therefore again points towards an $SO(10)$ theory where
such high scales are naturally generated.

To confront $SO(10)$ models with the solar neutrino data, one must make
precise predictions of the neutrino masses and mixing angles.  This
requires, however, detailed
information of the Dirac neutrino mass matrix as well as the Majorana
matrix.  In grand unified theories (GUTs), it is possible to
relate the
quark masses with the lepton masses.  It used to be
thought that in simple  $SO(10)$ models,
the charge $-1/3$ quark mass matrix is
related to the charged
lepton matrix and
the neutrino Dirac
mass matrix is related
to the charge $2/3$ quark matrix at the unification scale.
No simple
simple way was known, in general, to relate
the heavy Majorana matrix to the
charged fermion observables.  This prevented any prediction of
light neutrino spectrum without making extra symmetry assumptions.

In a recent paper,[15] it was shown Babu and this author that
this situation resulted from an incomplete analysis of the
Higgs sector of the model and we showed that in a class of minimal $SO(10)$
models,  not only the Dirac neutrino matrix, but the Majorana
matrix also gets related to observables in the charged fermion sector.
This leads to a complete prediction for the neutrino masses and
mixings without adding any extra Higgs multiplets or any new
symmetries to the theory.
We use a simple Higgs system with
one (complex) {\bf 10}
and one {\bf 126} that have Yukawa couplings to fermions.
The {\bf 10} is needed for
quark and lepton masses, the
{\bf 126} is needed for the see--saw mechanism.  Crucial to the
predictivity of the neutrino spectrum is the observation that
the standard model doublet contained in the {\bf 126}
receives an induced vacuum expectation
value (vev) at tree--level.  In its absence, one would
have the asymptotic mass
relations $m_b=m_\tau,~m_s=m_\mu,~m_d=m_e$.
While the first relation would lead to
a successful prediction of $m_b$ at low energies, the last two
are in disagreement with observations.  The induced vev of the standard
doublet of {\bf 126} corrects these bad relations and at the
same time
also relates the Majorana neutrino mass matrix to
observables in the charged fermion sector, leading to a predictive
neutrino spectrum.

\bigskip
\noindent{\elevenbf 2. Minimal SO(10) GUT without supersymmetry:}
\bigskip

In this section,
we shall consider non--SUSY $SO(10)$ model to illustrate our
mechanism. In the minimal scenerio defined here, the $SO(10)$
symmetry
 breaks to the standard model via
the $SU(2)_L \times SU(2)_R
\times SU(4)_C \equiv G_{224}$ chain.
The breaking of $SO(10)$ via $G_{224}$ is achieved by either
a {\bf 54} or
a {\bf 210} of Higgs.  The {\bf 210} also breaks
the discrete $D$--parity,[16] whereas the {\bf 54} preserves it.
$D$--parity is a local discrete
$Z_2$--subgroup of $SO(10)$, under $D$, a fermion field $f$ transforms
into its charge conjugate $f^c$.  Breaking of $D$--parity at the GUT
scale makes the see--saw mechanism natural[17].It further changes the
evolution of the gauge coupling constants by making the Higgs
spectrum left-right asymmetric. It also eliminates the cosmological
domain wall problem that can arise if this $Z_2$ local symmetry
 survived to the intermediate scale.
  We will therefore
work with the {\bf 210} Higgs multiplet, although our mechanism to
cure the fermion mass problem and predict the neutrino masses is
independent of this choice. We will denote this  chain with
$SU(4)_c$ as intermediate symmetry as chain A.
 If instead of a $\bf 210$ dim. Higgs
multiplet, we chose a $\bf 45$ plus $\bf 54$ combination, then the
$SO(10)$ symmetry would break down to $SU(2)_L\times SU(2)_R\times
U(1)_{B-L}\times SU(3)_c\equiv G_{2213}$ group.In this case, also
D-parity breaks at the GUT scale , thereby making the see-saw
formula natural.
 We will denote
this case as chain B.
The second stage of symmetry breaking
goes via the {\bf 126} in both cases.
  Finally, the standard model electro--weak symmetry breaking
proceeds via the {\bf 10}.

\bigskip
{\it 2a. Mass scales and Proton decay:}
\bigskip

Before discussing the fermion sector of the theory, let us discuss
the mass scales as predicted by the low energy values of $ sin^2\theta_W$,
$\alpha_{strong}$ and $\alpha_{em}$. This analysis was first carried
out for chains with D-parity breaking by Chang et. al. in ref.13 and
recently been reanalyzed by Deshpande et al.[13].
 The threshold corrections
due to heavy particles was carried out
recently by Parida and this author[18].
In the threshold correction analysis, the survival hypothesis was
used to determine which Higgs submultiplets are at what mass scale
and then allowing for reasonable uncertainties in the heavy Higgs
boson masses, the uncertainties in the mass scales was estimated.
It was found that the different mass scales and the uncertainties
in their values depend on the symmetry breaking chain and we have:

\noindent {Chain A:}
\begin{eqnarray}
M_U = 10^{15.8^{+.8}_{-1.7}\pm.2} GeV~~\nonumber \\
M_I = 10^{11.5^{+2.8}_{-1.5}\pm.02} GeV
\end{eqnarray}

\noindent {Chain B:}
\begin{eqnarray}
M_U = 10^{15.8\pm.1\pm.25}GeV~~\nonumber  \\
M_I = 10^{9^{+.6}_{-.3}\pm.18}GeV
\end{eqnarray}

 The significance of these results is that the $M_U$ governs the
life--time of the proton in $SO(10)$ models whereas the $M_I$
governs the neutrino masses vis the see--saw mechanism. We will come
to the neutrino masses later. The proton life--time in both symmetry
breaking chains is given by[18]:

\begin{equation}
Chain A:~~~~~~~~\tau_p = 1.6\times 10^{35\pm.7\pm .9^{+3.2}_{-6.8}}years\\
\end{equation}
\begin{equation}
chain B:~~~~~~~~\tau_p = 1.6\times 10^{35\pm.7\pm1.0\pm.8}years
\end{equation}

In the above formulae, the last uncertainties are from the threshold
corrections whereas the the first and second are from the
matrix element and $\alpha_{str}$ uncertainties respectively.
It is clear from eq.(3) that the threshold uncertainties in $M_I$
and $M_U$ are so large that one could consider this as an almost
single scale theory. This possibility has been studied in detail
in ref.19 and indeed, the results turn out to be quite consitent
with the present lower limits on proton lifetime.

\bigskip
{\it 2b. Fermion masses:}
\bigskip

Let us now turn to the fermion sector. we denote
the three families belonging to
{\bf 16}--dimensional spinor representation of $SO(10)$ by
$\psi_a$, $a=1-3$, the complex {\bf 10}--plet of Higgs by $H$, and
the {\bf 126}--plet of Higgs by $\Delta$,
the Yukawa couplings can be written down as
\begin{equation}
L_Y = h_{ab}\psi_a\psi_bH + f_{ab}\psi_a\psi_b\overline{\Delta} + H.C.
\end{equation}
Note that since the {\bf 10}--plet is complex, one other coupling
$\psi_a\psi_b\overline{H}$ is allowed in general.  In SUSY--$SO(10)$, the
requirement of supersymmetry prevents such a term.  In the non--SUSY
case, which we are focussing on in this section,
 we forbid this term by imposing a $U(1)_{PQ}$
symmetry, which may anyway be needed in order to solve the strong CP
problem.

The {\bf 10} and {\bf 126} of Higgs have the following decomposition
under $G_{224}$:
${\bf 126} \rightarrow (1,1,6)+(1,3,10)+(3,1,\overline{10})+(2,2,15)$,
${\bf 10} \rightarrow (1,1,6)+(2,2,1)$.
Denote the $(1,3,10)$ and $(2,2,15)$ components of
$\Delta({\bf 126})$
by $\Delta_R$ and
$\Sigma$ respectively and the $(2,2,1)$ component of $H({\bf 10})$ by
$\Phi$.
The vev $<\Delta_R^0> \equiv v_R \sim 10^{12}~GeV$
breaks the intermediate symmetry down to
the standard model and generates Majorana
neutrino masses given by $fv_R$.
$\Phi$  contains two standard model doublets
which acquire
vev's denoted by $\kappa_u$ and $\kappa_d$ with
$\kappa_{u,d} \sim 10^{2}~GeV$.
$\kappa_u$ generates charge 2/3 quark as well as Dirac neutrino
masses, while $\kappa_d$ gives rise to $-1/3$ quark and charged lepton
masses.

Within this minimal picture, if $\kappa_u,~\kappa_d$ and $v_R$ are
the only vev's
contributing to fermion masses, in addition to
the $SU(5)$ relations $m_b=m_\tau,~
m_s=m_\mu,~m_d=m_e$, eq. (7) will also lead to the unacceptable relations
$m_u:m_c:m_t = m_d:m_s:m_b$.
Moreover, the
Cabibbo-Kobayashi-Maskawa (CKM) mixing matrix will be identity.
Fortunately, within this minimal scheme, we have found
new contributions
to the fermion mass matrices which are of the right order of
magnitude to correct these bad relations.  To see this, note that
the scalar potential contains, among other terms, a crucial term
\begin{equation}
V_1 =\lambda \Delta \overline{\Delta} \Delta H +H.C.
\end{equation}
Such a term is invariant under the $U(1)_{PQ}$ symmetry.  It will be
present in the SUSY $SO(10)$ as well, arising from the {\bf 210}
$F$--term or perhaps as a Planck induced term etc.
This term induces vev's for the standard doublets contained in the $\Sigma$
multiplet of {\bf 126}.  The vev arises through a term
$\overline{\Delta}_R\Delta_R \Sigma \Phi$ contained in $V_1$.

We can estimate the magnitudes of the induced vev's
of $\Sigma$ (denoted by $v_u$ and $v_d$ along the up
and down directions) assuming the survival hypothesis to hold:
\begin{equation}
v_{u,d} \sim \lambda \left({v_R^2}\over {M_{\Sigma_{u,d}}^2}\right)
\kappa_{u,d}~~.
\end{equation}
Suppose $M_U \sim 10^{15}~GeV$, $M_I \sim 3 \times 10^{12}~GeV$
as in the chain A
and $M_{\Sigma} \sim 10^{14}~
GeV$, consistent with survival hypothesis, then $v_u$ and
$v_d$ are of order 100 MeV, in the right range for correcting the bad
mass relations.  We emphasize that there is no need for a
second fine--tuning to generate such induced vev's.  In the SUSY
version with no intermediate scale, the
factor $(v_R^2/M_{\Sigma}^2)$ is not a suppression, so
the induced vev's
can be as large as $\kappa_{u,d}$. This is the key observation
of ref.15, which leads to the predictive $SO(10)$ model in the
neutrino sector, while removing the disagreement with charged
fermion, masses that were thought to exist in the minimal model before.

Another way to view this is to realize that the effect of the
 mixing is to leave a pair of light standard model doublets
are admixtures of doublets in $(2,2,1)$ and $(2,2,15)$ multiplets
and these light doublets acquire vevs at low energies leading to
the same mass patterns as before.

It is also worth noting that if we considered the
symmetry breaking chain B, then since the intermediate scale
is at most $10^{10}$GeV, the induced vevs $v_u$ and $v_d$
would be very small and would not provide any useful
correction to the bad charged fermion mass relations. Thus,
our scenerio would prefer the $SU(4)_c$ intermediate symmetry
chain.

\bigskip
\noindent{\elevenbf 3. The charged fermion masses:}
\bigskip

We are now in a position to write down the quark and lepton mass
matrices of the model:
\begin{eqnarray}
M_u = h \kappa_u +fv_u~~&~&~~M_d = h \kappa_d+f v_d \nonumber \\
M_{\nu}^D = h \kappa_u-3 f v_u~~&~&~~M_l=h \kappa_d-3fv_d\nonumber \\
M_{\nu}^M &=& f v_R~.
\end{eqnarray}
Here $M_{\nu}^D$ is the Dirac neutrino matrix and $M_{\nu}^M$ is the
Majorana mass matrix.

Before proceeding, we should specify the origin of CP violation in the
model.
We shall assume that it is spontaneous or soft, that will keep the
number of parameters at a minimum.
The Higgs sector described above already has enough structure to
generate realistic CP violation either softly or spontaneously.
The Yukawa
coupling matrices $h$ and $f$ in this case
are real and symmetric.  Although there
will be three different phases in the vev's (one common phase for
$\kappa_u$ and $\kappa_d$ and one each for $v_u$ and $v_d$),
only two combinations enter into the mass matrices,
as the overall phase can be removed from
each sector.  We shall bring these two phases into $v_u$ and $v_d$
and hence
forth denote them by $v_ue^{i \alpha}$ and $v_d e^{i \beta}$.

To see the predictive power of the model as regards the neutrino
spectrum, note that we can choose a basis
where one
of the coupling matrices, say $h$, is real and diagonal.  Then there are
13 parameters in all, not counting the superheavy scale $v_R$: 3
diagonal elements of the matrix $h \kappa_u$, 6 elements of $f v_u$, 2
ratios of vev's $r_1=\kappa_d/\kappa_u$ and $r_2=v_d/v_u$, and the two
phases $\alpha$ and $\beta$.
These 13 parameters are related to the 13 observables in the
charged fermion sector, viz., 9 fermion masses, 3 quark mixing
angles and one CP violating phase.  The light neutrino mass matrix will
then be completely specified in terms of other physical observables
and the overall scale $v_R$.  That would lead to 8
predictions in the lepton sector:  3 leptonic
mixing angles, 2 neutrino mass ratios and 3 leptonic CP violating
phases.

The relations of eq. (10) hold at the intermediate
scale $M_I$ where quark--lepton
symmetry and left--right symmetry are intact.  There are calculable
renormalization corrections to these relations below $M_I$.  The quark and
charged lepton masses as well as the CKM matrix elements run between $M_I$ and
low energies.  The neutrino masses and mixing angles, however, do not
run below $M_I$, since the right-handed neutrinos have masses of order
$M_I$ and decouple below that scale.  The predictions in the neutrino
sector should then be arrived at by first extrapolating the charged fermion
observables to $M_I$.

We fix
the intermediate scale at $M_I = 10^{12}~GeV$ and use the one--loop
standard model renormalization group equations to track the running of
the gauge couplings between $M_Z$ and $M_I$.

To compute the renormalization factors, we choose as low energy inputs
the gauge couplings at $M_Z$ to be
$\alpha_1 (M_Z) = 0.01688,~\alpha_2 (M_Z) = 0.03322,~\alpha_3 (M_Z)
=0.11$.
For the light quark (running) masses, we choose values listed in Ref.
20.  The top--quark mass will be allowed to vary between 100 and 200 GeV.
Between 1 GeV and $M_Z$, we use two--loop QCD renormalization group
equations for the running of the quark masses and the $SU(3)_C$
gauge coupling,[20] treating
particle thresholds as step functions.  From $M_Z$ to $M_I$, the
running factors are computed semi--analytically both for the fermion
masses and for the CKM angles
by using the
one--loop renormalization group equations for the Yukawa couplings
and keeping the heavy top--quark contribution[21].  The running factors,
defined as $\eta_i = m_i(M_I)/m_i(m_i)$
[$\eta_i=m_i(M_I)/m_i(1~GeV)$
for light quarks $(u,d,s)$] are $\eta (u,c,t)=(0.273,0.286, 0.506)$,
$\eta (d,s,b)=$ $(0.279,0.279,0.327)$,
$\eta (e,\mu,\tau)=0.960$ for the case of
$m_t=150~GeV$.
The (common) running factors for
the CKM angles (we follow the parameterization advocated by the Particle
Data Group)
$S_{23}$ and $S_{13}$ is 1.081 for $m_t=150~GeV$.
The Cabibbo angle $S_{12}$ and the KM phase
$\delta_{KM}$ are essentially unaltered.

Let us first analyze the mass matrices of eq. (10) in the limit of CP
conservation.  We shall treat spontaneous CP violation arising through
the phases of the vev's $v_ue^{i \alpha}$ and $v_de^{i \beta}$ as small
perturbations.  This procedure will be justified a posteriori.  In fact,
we find that realistic fermion masses, in particular the first family
masses, require these phases to be small.

We can rewrite the mass matrices $M_l, M_\nu^D$ and $M_{\nu}^M$ of eq.
(10) in terms of the quark mass matrices and three ratios of
vev's -- $r_1=\kappa_d/\kappa_u,~r_2=v_d/v_u~,R=v_u/v_R$:
\begin{eqnarray}
M_l &=& {{4 r_1 r_2}\over{r_2-r_1}} M_u-{{r_1+3 r_2}\over
{r_2-r_1}} M_d ~~,\nonumber \\
M_\nu^D &=& {{3 r_1+r_2}\over {r_2-r_1}} M_u-{4 \over {r_2-r_1}}M_d ~~,
\nonumber \\
M_\nu^M &=& {1 \over R} {{r_1} \over {r_1-r_2}}
M_u -{1 \over R}{1 \over {r_1-r_2}}M_d~~.
\end{eqnarray}

It is convenient to go to a basis where $M_u$ is diagonal.  In that
basis, $M_d$ is given by $M_d=V M_d^{diagonal} V^T$, where
$M_d^{diagonal}={\rm diagonal}(m_d,m_s,m_b)$ and $V$ is the CKM matrix.  One
sees that $M_l$ of eq. (11) contains only physical observables from the
quark sector and two parameters $r_1$ and $r_2$.  In the CP--conserving
limit then, the three eigen--values of $M_l$ will lead to one mass prediction
for the charged fermions.  To
see this prediction, $M_l$ needs to be diagonalized.  Note first that by
taking the Trace of $M_l$ of eq. (11), one obtains a
relation for $r_1$ in terms of
$r_2$ and the charged fermion masses.  This is approximately
\begin{equation}
r_1 \simeq \left(m_\tau+3 m_b\right)/4 m_t
\end{equation}
(as long as
$r_2$ is larger than $m_b/m_t$).
Since $|m_b| \simeq |m_\tau|$ at the
intermediate scale to within 30\% or so, depending on the relative sign
of $m_b$ and $m_\tau$, $r_1$ will be close to either $m_b/m_t$ or
to $(m_b/2m_t)$.
Note also that if $r_2 \gg r_1$, $M_l$ becomes
independent of $r_2$, while $M_{\nu}^D$ retains some dependence:
\begin{equation}
M_l \simeq 4 r_1 M_u -3 M_d,~
M_{\nu}^D \simeq M_u-{4 \over {r_2}}M_d.
\end{equation}
This means that the parameter $r_2$ will only be
loosely constrained from the charged fermion sector.

We do the fitting as follows.
For a fixed value of $r_2$, we determine $r_1$ from the Tr$(M_l)$
using the input
values of the masses and the renormalization factors
discussed above.  $M_l$ is then diagonalized
numerically.  There will be two mass relations among charged fermions.
Since the charged lepton masses are precisely known at low energies,
we invert these relations to predict the $d$--quark and $s$--quark
masses.  The $s$--quark mass is sensitive to the muon
mass, the $d$--mass is related to the electron mass.
This procedure is repeated for other values of
$r_2$.  For each choice, the light neutrino masses and the leptonic CKM
matrix elements are then computed using the see--saw formula.

\bigskip
\noindent{\elevenbf 4. Predictions for neutrino masses and mixings:}
\bigskip

We find that there are essentially three different solutions.  A
two--fold ambiguity arises
from the unknown relative sign of $m_b$ and $m_\tau$ at $M_I$.
Although solutions exist for both signs, we have found that a relative
minus sign tends to result in somewhat large value of $m_s/m_d$.
Our numerical fit shows that the loosely constrained parameter
$r_2$ cannot be
smaller than 0.1 or so, otherwise the $d$--quark mass comes out
too small.  Now, the light neutrino spectrum is sensitive to
$r_2$ only when $r_2 \sim 4 m_s/m_c\sim \pm 0.4$, since the two terms in
$M_\nu^D$
become comparable (for the second family) then.  Two qualitatively
different solutions are
obtained depending on whether $r_2$ is near $\pm 0.4$ or not.

Numerical results for the three different cases are presented below.
The input values of the CKM mixing angles are chosen for all cases to be
$S_{12}=-0.22,~S_{23}=0.052,~S_{13}=6.24 \times 10^{-3}$.
Since $\delta_{KM}$ has been set to zero for now, we have allowed for
the mixing
angles to have either sign.  Not all signs result in acceptable
quark masses though.  Similarly, the fermion masses can have either sign,
but these are also restricted.
The most stringent constraint comes from the $d$--quark
mass, which has a tendency to come out too small.  Acceptable solutions are
obtained when $\theta_{23},~\theta_{13}$ are
in the first quadrant and $\theta_{12}$ in the fourth quadrant.

Solution 1:
\begin{eqnarray}
{\rm Input}:  m_u(1~GeV) &=& 3~MeV,~~m_c(m_c)=1.22~GeV,~~m_t =
150~GeV \nonumber \\
m_b(m_b)&=&-4.35~GeV,~~r_1=-1/51.2,~~r_2=2.0 \nonumber \\
{\rm Output}:  m_d(1~GeV) &=& 6.5~ MeV,~~m_s(1 GeV)=146~MeV \nonumber \\
\left(m_{\nu_e},~m_{\nu_{\mu}},~m_{\nu_\tau}\right) &=& R\left(2.0 \times
10^{-2},9.9,-2.3 \times 10^4\right)~GeV \nonumber \\
V_{KM}^{\rm lepton} &=& \left(\matrix{0.9488 & 0.3157 & 0.0136 \cr
-0.3086 & 0.9349 & -0.1755 \cr -0.0681 & 0.1623 & 0.9844}\right)~~.
\end{eqnarray}

Solution 2:
\begin{eqnarray}
{\rm Input}: m_u(1~GeV)&=& 3~MeV,~~m_c(m_c)=1.22~GeV,~~m_t=150~GeV\nonumber
\\
m_b(m_b) &=& -4.35~GeV,~~r_1=-1/51,~~r_2=0.2 \nonumber \\
{\rm Output}: m_d(1~GeV) &=& 5.6~MeV,~~ m_s(1~GeV)=156~MeV \nonumber \\
\left(m_{\nu_e},m_{\nu_\mu},m_{\nu_\tau}\right) &=& R\left(7.5 \times
10^{-3},2.0,-2.8 \times 10^3\right)~GeV \nonumber \\
V_{KM}^{\rm lepton} &=& \left(\matrix{0.9961 & 0.0572 & -0.0676 \cr
-0.0665 & 0.9873 & -0.1446 \cr 0.0584 & 0.1485 & 0.9872}\right)~~.
\end{eqnarray}

Solution 3:
\begin{eqnarray}
{\rm Input:} m_u(1~GeV) &=& 3~MeV,~m_c(m_c) = 1.27~GeV,~m_t=150~GeV
\nonumber \\
m_b(m_b) &=& -4.35~GeV,~~r_1=-1/51.1,~~r_2=0.4 \nonumber \\
{\rm Output}: m_d(1~GeV) &=& 6.1~MeV,~~ m_s(1~GeV)=150~ MeV \nonumber \\
\left(m_{\nu_e},m_{\nu_\mu},m_{\nu_\tau}\right) &=& R\left(4.7 \times
10^{-2},1.4,-5.0 \times 10^3\right)~GeV \nonumber \\
V_{KM}^{\rm lepton} &=& \left(\matrix{0.9966 & 0.0627 & -0.0541 \cr
-0.0534 & 0.9858 & 0.1589 \cr 0.0633 & -0.1555 & 0.9858}\right)~~.
\end{eqnarray}

Solution 1 corresponds to choosing $r_1 \sim m_b/m_t$.  All the charged
lepton masses are negative in this case.  Since $r_2$ is large, the Dirac
neutrino matrix is essentially $M_u$, which is diagonal; so is the
Majorana matrix.  All the
leptonic mixing angles arise from the charged lepton sector.  Note that
the predictions for $m_d$ and $m_s$ are within the range quoted
in Ref. 20.  The ratio $m_s/m_d = 22$ is within the allowed range from
chiral perturbation theory estimates[22].
The mixing angle sin$\theta_{\nu_e-\nu_\mu}$ relevant for solar
neutrinos is 0.30, close to the Cabibbo angle.  Such a value may already
be excluded by a combination of all solar neutrino data taken at the 90\% CL
(but not at the 95\% CL)[7].  Actually, within the model, there is a
more stringent constraint.  Note that the $\nu_\mu-\nu_\tau$ mixing
angle is large, it is approximately $3|V_{cb}| \simeq 0.16$.  For that
large a mixing, constraints from $\nu_\mu-\nu_\tau$
oscillation experiments imply[23] that $|m_{\nu_\tau}^2-
m_{\nu_{\mu}}^2| \le 4~eV^2$.  Solution 1 also has
$m_{\nu_\tau}/m_{\nu_\mu} \simeq 2.3 \times 10^3$, requiring that
$m_{\nu_{\mu}} \le 0.9 \times 10^{-3}~eV$.  This is a factor of 2 too
small for $\nu_e-\nu_\mu$ MSW oscillation for the
solar puzzle (at the 90\% CL), but perhaps is not excluded completely,
once astrophysical uncertainties are folded in.
If $\nu_\tau$ mass is around $2 \times 10^{-3}~eV$,
$\nu_e-\nu_{\tau}$ oscillation may be relevant, that mixing angle is
$\simeq 3|V_{td}| \simeq 6\%$.  It would
require the parameter $R=v_u/v_R \sim 10^{-16}$ or $v_R \sim 10^{16}~
GeV$ for $v_u \sim 1~GeV$.  Such a scenario fits very well within
SUSY--$SO(10)$.

Solution 2 differs from 1 in that $r_2$ is smaller,
$r_2=0.2$.  The ratio $m_s/m_d =27.8$ is slightly above the limit in
Ref. 22.  The $1-2$ mixing in the
neutrino sector is large in this case, so it can cancel the
Cabibbo like mixing arising from the charged lepton sector.  As we vary
$r_2$ from around 0.2 to 0.6, this cancellation becomes stronger, the
$\nu_e-\nu_\mu$ mixing angle becoming zero for a critical value of
$r_2$.  For larger $r_2$, the solution will approach Solution 1.
The
$\nu_\mu-\nu_\tau$ mixing angle is still near $3 |V_{cb}|$, so as
before,  $m_{\nu_{\tau}} \le 2~ eV$.  From the $\nu_\tau/\nu_\mu$ mass
ratio, which is $1.4 \times 10^3$ in this case, we see that
$m_{\nu_\mu} \le 1.5 \times 10^{-3}~eV$.  This is just within the
allowed range[7] (at 95\% CL) for small angle non--adiabatic
$\nu_e-\nu_\mu$ MSW oscillation, with a predicted count rate of about
50 SNU for the Gallium experiment.
Note that there is a lower limit of about 1 eV for
the $\nu_\tau$ mass in this case.  Forthcoming experiments should then
be able to observe $\nu_\mu-\nu_\tau$ oscillations.  A $\nu_\tau$ mass
in the (1 to 2) eV range can also be cosmologically significant, it can
be at least part of the hot dark matter.
In SUSY $SO(10)$, $\nu_e-\nu_\tau$ oscillation (the relevant mixing is
about $3|V_{td}| \simeq 5\%$), could account for the solar neutrino puzzle.

Solution 3 corresponds to choosing $r_1 =0.4$.  $m_s/m_d=24.6$ is within
the allowed range.
However, the mass
ratio $\nu_\tau/\nu_\mu$ is $\sim 3.6 \times 10^3$, and
sin$\theta_{\mu \tau} \simeq 3 |V_{cb}|$
so $\nu_e-\nu_\mu$ oscillation
cannot be responsible for solar MSW.  As in other cases,
$\nu_e-\nu_\tau$ MSW oscillation with a 6\% mixing is a viable possibility.

Let us now re-instate the CP--violating
phases $\alpha$ and $\beta$ in the vev's perturbatively.  Small
values of the phases are sufficient to account for realistic CP violation
in the quark sector.
We shall present details for the
case of Solution 2 only, others are similar.  We also
tried to fit all the charged
fermion masses and mixing angles for large phases, but found no
consistent solution.

First we make a basis transformation to go from the basis where $M_u$
is diagonal to one where the matrix
$h \kappa_u$ is diagonal.  It is easier to introduce phases in that
basis.
For $\alpha=3.5^0,~\beta=4.5^0$, the
CP--violating parameter $J$ for the quark system[24] is $J \simeq 1 \times
10^{-5}$, which is sufficient to accommodate $\epsilon$ in the
neutral $K$ system.  The leptonic CP violating phases are
correspondingly small, for eg., the analog of $J$ is
$J_l \simeq 7 \times 10^{-5}$.
These small phases modify the first family masses
slightly, but the effect is less than 10\%.  Our predictions for the neutrino
mixing angles are essentially unaltered.

Recently, three more solutions for neutrino masses have been found by
Lavoura, which give good fits to the charged fermion masses[25]. We refer
to ref.25 for details on them.

\bigskip
\noindent{\elevenbf 5. Supersymmetric minimal SO(10) and its implications:}
\bigskip

        In this section, we consider the supersymmetric version of
our model. The fermions are part of the chiral superfields transforming
as the ${16}$--dimensional spinor representation of the $SO(10)$ group and
are denoted by $\psi$. We will choose the Higgs multiplets for the
supersymmetric version  the same way for the non-SUSY case,
i.e., Higgs fields belonging to $\bf 10$, ${\bf \overline{126}}$ and $\bf 210$
representations (denoted by $H$, $\bar{\Delta}$ and $\Phi$ respectively)
but we have to include a $\bf 126$ dimensional multiplet (denoted by $\Delta$)
to maintain supersymmetry down to the electro-weak scale. The
most general superpotential for the model can be written as:
\begin{equation}
W= W_m
+\mu_1 \Phi^2 + \lambda_1 \Phi^3 + \lambda_2\Delta \bar{\Delta}\Phi
+\mu_2 \Delta \bar{\Delta} + \lambda_3 \Phi \Delta H + \lambda_4
\Phi \bar{\Delta} H + \mu_3 H^2
\end{equation}
where
\begin{equation}
 W_m=h_{ab}\psi_a \psi_b H + f_{ab}\psi_a \psi_b \bar{\Delta}~~.
\end{equation}
   The gauge and supersymmetry breaking is determined by writing down
the $F$-terms and setting them to zero at the GUT scale. The detailed
analysis of this is given in ref. 26. Here, we simply quote the
result of that paper which says that if we do not fine--tune any of the
parameters of the superpotential in the above equation, the vanishing
of $F$-terms requires that the $SO(10)$ symmetry break down to the
standard model in one step. This implies that $v_R\simeq M_U$. This has
several interesting implications as we will see later. Unfortunately,
as in all SUSY GUT models, in the supersymmetric limit, there are
several degenate vacua corresponding to the symmetry groups i) $SU(5)$,
ii) $SU(2)_L\times SU(2)_R\times U(1)_{B-L}\times SU(3)_c$;
iii) $SU(2)_L\times U(1)_{I_{3R}}\times U(1)_{B-L}\times SU(3)_c$
iv) $SU(5)\times U(1)$ v) SO(10).
 Once supergravity corrections are included
this degeneracy  disappears, and by appropriate choice
of parameters, we can always choose the global minimum to correspond
to the standard model.

\bigskip
\noindent{\it Doublet-Triplet Splitting:}
\bigskip

  The second point to notice is that in the Higgs potential derived
from this superpotential, there are terms of the form $\Delta
\bar{\Delta}\Delta H$ proportional to $\lambda_2 \lambda_3$ and
$\lambda_2 \lambda_4$ so that a vev for the $\bf (2,2,15)$
sub-multiplet of $\bf 126$ is induced once $SO(10)$ and $B-L$
are broken, in analogy with the non--SUSY case. The detailed analysis of
this question is tied to the question of doublet-triplet splitting
of SUSY-GUT models. Let us therefore briefly discuss this question
below. In the minimal $SO(10)$ model there are eight light Higgs
doublets, four $H_u$ type (i.e. capable of giving masses to up
quarks ) and four $H_d$ type which can give mass to down type quarks.
After GUT symmetry breaking , one therefore has a $4\times 4$ matrix
and for the theory to reproduce low energy physics, this matrix must
have two massless eigen-modes. In general this will require some
fine tuning of parameters so that all doublets do not
acquire GUT scale mass. At the same time, this model has a
$5\times 5$ mass matrix involving the color triplet fields. Successful
doublet--triplet splitting requires that the same fine--tuning
that leaves a pair $(H_u,H_d)$  massless should not at the same time
leave any triplets below the GUT scale, otherwise, not only will
there be problem with proton life--time, there will also be problem
with unification of couplings. The doublet--triplet splitting is called
natural when the required fine tuning is guaranteed by some
symmetry.

To study this question in detail, let us isolate the various
doublet and triplet fields in the model : The four up Higgs
doublets are denoted by: $H_u$ from $\bf 10$, $\sigma_u$ and
$\overline{\sigma}_u$ from $\bf 126$ and $\overline {\bf 126}$
respectively and $\alpha_u$ from $\bf 210$ and similarly four
down type doublets from the same Higgs representations.
The five color triplets fields belonging to $\bf 3$ rep. and
electric charge $-1/3$ are denoted by $\zeta_1$ from $\bf 10$,
 $\overline{\gamma}_1$ from $\overline{\bf 126}$, $\gamma_2$
from $\bf 126$, $\omega$ from $\bf (1,3,15)$ in $\bf 210$
and $\overline{\delta}_R$ from $\bf (1,3,10)$ in $\overline{\bf 126}$.
There is of course a similar set belonging to $\overline{\bf 3}$
to be denoted by $\zeta_2$,$\overline{\gamma}_2$, $\gamma_1$,
$\overline{\omega}$ and $\delta_R$. The $4\times 4$ Dirac matrix
involving the doublets can be written as follows: ( the basis for this
matrix is ($ H_u, \sigma_u, \overline{\sigma}_u, \alpha_u$) denoting the
columns and corresponding fields with subscripts d denoting the
rows).

\begin{equation} M_D = \left( \begin{array}{cccc}
 \mu_3   &  \lambda_3 v_U  & \lambda_4 v_U  &  \lambda_4 v_R \\
\lambda_3 v_U  &  0  &  \tilde{\mu}_2  &  \lambda_2 v_R \\
\lambda_4 v_U  &  \tilde{\mu}_2  &  0  &  0 \\
\lambda_3 v_R  &  0  &  \lambda_2 v_R & \tilde{\mu}_1  \end{array} \right)
\end{equation}

In the above matrix, $v_R$ and $v_U$ stand for the vev's of the
$\overline{\bf 126}$ and $\bf 210$ dim. Higgs multiplets respectively
and the symbols $\tilde{\mu}_i$ denote some combination of $\mu_i$
with the vev contributions. The color triplet matrix in the basis
$(\zeta_1,\overline{\gamma}_1,\gamma_2,\overline{\omega} , \delta_R)$
for columns and corresponding $\overline{\bf 3}$ denoting rows is:

\begin{equation} M_T= \left( \begin{array}{ccccc}
\mu_3 & \lambda_4 v_U & \lambda_3 v_U & \lambda_4 v_R & \lambda_4 v_U \\
\lambda_4 v_U & 0  &  \tilde{\mu}_2 &  0  &  0  \\
\lambda_3 v_U  &  \tilde{\mu}_2  &  0  &  \lambda_2 v_R  & \lambda_2 v_U \\
\lambda_3 v_R  & \lambda_2 v_R &  0  &  \tilde{\mu}_1 & \lambda_2 v_R \\
\lambda_3 v_U  &  \lambda_2 v_U  & 0  & \lambda_2 v_R & \tilde{\mu}_2
\end{array}  \right)
\end{equation}

We wish to point out that, the entries in the above matrices
do not contain the detailed numerical coefficients nor the
various vevs corresponding to the multiplets- rather just the
orders of magnitudes.
It is clear that, one needs fine tuning to get a light pair of Higgs
doublets. For our scheme to work, we need a linear combination of
$H_u$ and $\overline{\sigma}_u$ (and similarly with subscript d)
to remain massless at the GUT scale. There are several ways to achieve
this. One particularly simple way is to set $ \lambda_3 =
\lambda_4 = 0 $ and add to the theory a Planck scale induced term
of the form $\bf 10$ $\bf 126$ $\overline{\bf 126}$ $\bf 10$/$M_{Pl}$
such that it balances against the $\mu_3$ term after GUT scale symmetry
breaking. At this stage, there are two light doublets, which arise
only from the $\bf 10$ Higgs. But if we add a further Planck induced
term of the form $\bf 10$ $\bf 126$ $\overline{\bf 126}$ $\bf 126$/$M_{Pl}$,
then this causes an admixture of $\overline{\sigma}_u$ with $H_u$
and similarly for $H_d$ as required. The corresponding color triplet
will however be somewhat light and considerations of proton decay
then suggests that the coefficients of the Planck induced term be
rather large.

  The pattern of the predictions for the neutrino masses and mixings
in the SUSY version is very similar to the non-SUSY version.

\bigskip
\noindent{\elevenbf 6. Summary and Conclusion:}
\bigskip

In summary, we feel that if the present indications for neutrino masses
in the micro-, milli-, and eV range continues to receive full confirmation,
then the SUSY $SO(10)$ may provide the correct framework for the
next stage in the search for further unification of forces and
matter. Our recent work described in this report would suggest
that the relevant $SO(10)$ theory could very well be the
minimal one with just three Higgs multiplets needed to break
the gauge symmetries appropriately. Just like the minimal $SU(5)$
had the unique feature that it predicted a value for both proton
lifetime and $sin^2\theta_W$, and could therefore be tested
experimentally, the  minimal $SO(10)$ grand unified
model leads to definite predictions for
 light neutrino masses and mixing angles
in terms of observables in the charged fermion sector and could
therefore be tested in the next round of neutrino oscillation
experiments. This in our opinion elevates minimal $SO(10)$ theory above
other competing GUT scenerios that can lead to nonvanishing
neutrino masses such as $SU(5)\times U(1)$ or $E_6$ theories
in terms of testability. In a sense, the status of minimal
$SO(10)$ is better than the minimal $SU(5)$ since here we fit
all charged fermion masses without invoking additional Higgs
multiplets. Of course if we want further predictability , say
in the charged fermion sector, we will certainly need new
symmetries. This line of research is currently being pursued.

 We also wish to point out that our approach
here has been orthogonal to some other recent attempts[27] based
on grand unification; we have kept the Higgs sector as simple as
possible and followed its consequences.  In particular, our results are
manifestly stable under radiative corrections.  Moreover, we have not
used any sort of family symmetry to make the model predictive.
Predictivity in the neutrino sector follows from minimality.
To repeat again, we have found three different
types of solutions for the neutrino spectrum.
In Solution 1, the $\nu_e-\nu_\mu$ mixing
angle is near the Cabibbo angle, while Solutions 2 and 3 have it much
smaller.  In all cases, $\nu_e-\nu_\tau$ mixing angle is predicted to be
near $3|V_{td}| \simeq 0.05$ and
$\nu_\mu-\nu_\tau$ mixing angle is
$\simeq 3 |V_{cb}| \simeq 0.15$ with the mass ratio
$m_{\nu_\tau}/m_{\nu_\mu} \ge 10^3$.  None of the solutions admit the
vacuum oscillation solution to the solar neutrino puzzle.
If it is
due to small angle non--adiabatic MSW, as in Solution 2,
$\nu_\mu-\nu_\tau$ oscillation should be observable in the
forthcoming experiments.

On the theoretical side,there is one important lesson for
$SO(10)$ model builders, i.e. whenever, there is a $\bf 10$
contributing to charged fermion masses
and $\overline{\bf 126}$ contributing to the right-handed
neutrino masses, one must include the induced vev contribution
also in the charged fermion sector. This will certainly effect
the predictivity of the models in the charged fermion sector.
Our model also has the potential to solve the strong CP problem:
note for instance that if $\lambda_4$ is set to zero, we have
a model with a softly broken PQ symmetry.

      The work of R.N.M. was supported by a grant from the
National Science Foundation . I would like to thank K.S.Babu
for collaboration in this work and Lee Dai Gyu for discussions.

\newpage
\noindent{\elevenbf 7. References}

\begin{enumerate}
\item J. Bahcall and M. H. Pinsonneault, {\it Rev. Mod. Phys.}
{\bf 64}, 885  (1993).\\
S.Turck-Chieze and I.Lopes, {\it Ap. J.} {\bf 408}, 347 (1993)
\item R. Davis et. al., in {\it Proceedings of the 21st International
Cosmic Ray Conference}, Vol. 12, ed. R.J. Protheroe (University of
Adelide Press, Adelide, 1990) p. 143; \\
K.S. Hirata et. al., {\it Phys. Rev. Lett.} {\bf 65}, 1297
(1990); \\
A.I. Abazov et. al., {\it Phys. Rev. Lett.} {\bf 67}, 3332 (1991); \\
P. Anselman et. al., {\it Phys. Lett.}
{\bf B285}, 376 (1992).
\item R.S.Raghavan et. al. {\it Phys. Rev.} {\bf D 44}, 3786 (1991).
\item R. Robertson, {\it Moriond Talk },(1993).
\item A.Acker, S. Pakvasa and J. Pantaleone, {\it Phys. Rev.} {\bf D 43}
,1754 (1991)\\
V.Barger, R.J.N.Phillips, {\it Phys. Rev. Lett.} {\bf 69}, 3135 (1992).
\item L. Wolfenstein, {\it Phys. Rev.} {\bf D 17}, 2369 (1978); \\
S.P. Mikheyev and A. Yu Smirnov, {\it Yad. Fiz.} {\bf 42}, 1441 (1985)
[{\it Sov. J. Nucl. Phys}, {\bf 42},
913 (1985)].
\item For recent updates after GALLEX results, see for example:\\
X. Shi, D.N. Schramm and J. Bahcall, {\it Phys. Rev. Lett.} {\bf 69},
717 (1992);\\
S.A. Bludman, N. Hata, D.C. Kennedy and P. Langacker, {\it Phys. Rev.}
{\bf D 47}, 2220 (1993).
\item K.Hirata et.al., {\it Phys. Lett.} {\bf B280}, 146 (1992)\\
D.W.Casper et. al., {\it Phys. Rev. Lett.} {\bf 66}, 2561 (1991)\\
Ch. Berger et. al., {\it Phys. Lett.} {\bf B227}, 489 (1989)\\
M. Aglietta et. al., {\it Europhys. Lett.} {\bf 8}, 611 (1989).
\item E.L.Wright et. al., {\it Ap. J.} {\bf 396}, L13 (1992)\\
R. Schaefer and Q. Shafi, BA-92-28 (1992)\\
J. A. Holtzman and J. Primack, {\it Ap. J.} {\bf 405}, 428 (1993).
\item W. Fratti et. al.,{\it Phys. Rev.} {\bf D 48}, 1140 (1993)\\
D. Perkins, {\it Nucl. Phys. } {\bf B399}, 1 (1993).
\item D. Caldwell and R. N. Mohapatra, {\it Phys. Rev.} {\bf D48},
3259 (1993).
\item M. Gell-Mann, P. Ramond and R. Slansky, in {\it Supergravity}, ed.
F. van Nieuwenhuizen and D. Freedman (North Holland, 1979), p. 315; \\
T. Yanagida, in {\it Proceedings of the Workshop on Unified Theory and
Baryon Number in the Universe}, ed. A. Sawada and H. Sugawara, (KEK,
Tsukuba, Japan, 1979);\\
R.N. Mohapatra and G. Senjanovic {\it Phys. Rev. Lett.} {\bf 44}, 912
(1980).
\item D. Chang, R.N. Mohapatra, M.K. Parida, J. Gipson and R.E. Marshak,
{\it Phys. Rev.} {\bf D 31}, 1718 (1985);\\
For a more recent analysis, see N.G. Deshpande, R. Keith and P.B. Pal,
Oregon Preprint OITS-484 (1992).
\item M. Fukugita and T. Yanagida, {\it Phys. Lett.} {\bf B174}, 45 (1986)\\
P. Langacker, R.D.Peccei and T. Yanagida,{\it Mod. Phys. Lett} {\bf A1},
541 (1986)\\
\item K. S. Babu and R. N. Mohapatra, {\it Phys. Rev. Lett.} {\bf 70},
2845 (1993).
\item D. Chang, R.N. Mohapatra and M.K. Parida, {\it Phys. Rev. Lett.}
{\bf 52}, 1072 (1982).
\item D. Chang and R.N. Mohapatra, {\it Phys. Rev.} {\bf D 32}, 1248
(1985).
\item R. N. Mohapatra and M. K. Parida, {\it Phys. Rev.} {\bf D 47}, 264
(1992).
\item L. Lavoura and L. Wolfenstein, {\it Carnegie- Mellon Univ. Preprint},
CMU-HEP93-03 (1993).
\item J. Gasser and H. Leutwyler, {\it Phys. Rept.} {\bf 87}, 77 (1982).
\item K.S. Babu, {\it Z. Phys.} {\bf C 35}, 69 (1987);\\
K. Sasaki, {\it Z. Phys.} {\bf C 32}, 149 (1986);\\
B. Grzadkowski, M. Lindner and S. Theisen, {\it Phys. Lett.} {\bf B198},
64 (1987).
\item D. Kaplan and A. Manohar, {\it Phys. Rev. Lett.} {\bf 56}, 2004 (1986).
\item N. Ushida et. al., {\it Phys. Rev. Lett.} {\bf 57}, 2897 (1986).
\item C. Jarlskog, {\it Phys. Rev. Lett.} {\bf 55}, 1039 (1985).
\item L. Lavoura, {\it Carnegie-Mellon Preprint}  (1993).
\item K. S. Babu and R. N. Mohapatra, in preparation.
\item J. Harvey, P. Ramond and D. Reiss, {\it Nucl. Phys.} {\bf B199},
223 (1982);\\
S. Dimopoulos, L. Hall and S. Raby, {\it Phys. Rev. Lett.} {\bf 68},
1984 (1992).\\
H. Arason et. al., {\it Phys. Rev.} {\bf D 47}, 232 (1992);\\
G. Lazarides and Q. Shafi, {\it Nucl. Phys.} {\bf B350}, 179 (1991)\\
H. Cheng, M. S. Gill and L. Hall, {\it LBL Preprint}  (1993).
\end{enumerate}

\end{document}